\documentclass[useAMS,usenatbib]{mn2e} 
\usepackage{color}
\usepackage{epsfig}
\usepackage{amsbsy}  
\usepackage{amssymb}

\newcommand{\be}{\begin{equation}} 
\newcommand{\ee}{\end{equation}}
\newcommand{\bea}{\begin{eqnarray}} 
\newcommand{\eea}{\end{eqnarray}}

\newcommand{\COSMOMC}{{\sc cosmomc}}

\def\lt{<}

 \def \lleq {\lower0.9ex\hbox{$\buildrel \lt \over \sim$} ~}

\title[A kinematical approach to dark energy studies] 
{A kinematical approach to dark energy studies}
\author[D.~Rapetti, 
S.~W.~Allen, M.~A.~Amin, R.~D.~Blandford] {David~Rapetti${}^{1}$\thanks{Email: drapetti@slac.stanford.edu},
  Steven~W.~Allen${}^{1}$, Mustafa~A.~Amin${}^{1}$, Roger~D.~Blandford${}^{1}$\\
  ${}^1$ Kavli Institute for Particle Astrophysics and Cosmology,
  Stanford University, 382 Via Pueblo Mall, Stanford 94305-4060, USA.}

\begin{document}
\date{Accepted ???, Received ???; in original form \today}
\pagerange{\pageref{firstpage}--\pageref{lastpage}} \pubyear{2005}

\maketitle
\label{firstpage}

\begin{abstract}
  We present and employ a new kinematical approach to cosmological
  `dark energy' studies. We construct models in terms of the
  dimensionless second and third derivatives of the scale factor
  $a(t)$ with respect to cosmic time $t$, namely the present-day value
  of the deceleration parameter $q_0$ and the cosmic jerk parameter,
  $j(t)$. An elegant feature of this parameterization is that all
  $\Lambda$CDM models have $j(t)=1$ (constant), which facilitates
  simple tests for departures from the $\Lambda$CDM paradigm. Applying
  our model to the three best available sets of redshift-independent
  distance measurements, from type Ia supernovae and X-ray cluster gas
  mass fraction measurements, we obtain clear statistical evidence for
  a late time transition from a decelerating to an accelerating phase.
  For a flat model with constant jerk, $j(t)=j$, we measure $q_{\rm
    0}=-0.81\pm 0.14$ and $j=2.16^{+0.81}_{-0.75}$, results that are
  consistent with $\Lambda$CDM at about the $1\sigma$ confidence
  level. A standard `dynamical' analysis of the same data, employing
  the Friedmann equations and modeling the dark energy as a fluid with
  an equation of state parameter, $w$ (constant), gives $\Omega_{\rm
    m}=0.306^{+0.042}_{-0.040}$ and $w=-1.15^{+0.14}_{-0.18}$, also
  consistent with $\Lambda$CDM at about the $1\sigma$ level. In
  comparison to dynamical analyses, the kinematical approach uses a
  different model set and employs a minimum of prior information,
  being independent of any particular gravity theory.  The results
  obtained with this new approach therefore provide important
  additional information and we argue that both kinematical and
  dynamical techniques should be employed in future dark energy
  studies, where possible. Our results provide further interesting
  support for the concordance $\Lambda$CDM paradigm.
\end{abstract}

\begin{keywords}
cosmology:observations -- cosmology:cosmological parameters -- 
cosmology:theory -- supernovae -- x-ray clusters -- dark energy
\end{keywords}

\section{Introduction}
\label{introduction}
                                     
The field of cosmology has made unprecedented progress during the past
decade. This has largely been driven by new observations, including
precise measurements of the spectrum of cosmic microwave background
(CMB) anisotropies \citep[][and references therein]{Spergel:03,
Spergel:06}, the distance-redshift relation to type Ia supernovae
\citep{Riess:98, Perlmutter:98, Knop:03, Riess04, Astier06}, the
distance-redshift relation to X-ray galaxy clusters \citep{Allen:02,
Allen:04, Ettori:03}, measurements of the mean matter density and
amplitude of matter fluctuations from X-ray clusters
\citep{Borgani:01, Allen:03, Reiprich:02, Schuecker:03, Voevodkin:04},
measurements of the matter power spectrum from galaxy redshift surveys
\citep{Tegmark:03,Cole:05, Eisenstein:05}, Lyman-$\alpha$ forest
studies \citep{Croft:00, Viel:04, Seljak:04} and weak lensing surveys
\citep{vanWaerbeke:00, Hoekstra:02, vanWaerbeke:05, Jarvis:05}, and
measurements of the Integrated Sachs-Wolfe effect \citep{Scranton:03,
Fosalba:03}.

These and other experiments have lead to the definition of the
so-called concordance $\Lambda$CDM cosmology. In this model, the
Universe is geometrically flat with only $\sim$4 per cent of the
current mass-energy budget consisting of normal baryonic matter.
Approximately 23 per cent is cold dark matter, which interacts only
weakly with normal baryonic matter but which clusters under the action
of gravity. The remaining $\sim 73$ per cent consists of smoothly
distributed quantum vacuum energy (the cosmological constant), which
pushes the Universe apart. This combination of matter and vacuum
energy leads to the expectation that the Universe should undergo a
late time transition from a decelerating to an accelerating phase of
expansion. Late-time acceleration of the Universe is now an observed
fact \citep[e.g.][]{Riess04, Allen:04, Astier06}. A transition from
a decelerating phase to a late-time accelerating phase is required to
explain both these late-time acceleration measurements and the
observed growth of structure.

Despite the observational success of the concordance $\Lambda$CDM
model, significant fine tuning problems exist. In particular,
difficulties arise in adjusting the density of the vacuum energy to be
a non-zero but tiny number, when compared with the value predicted by
standard theoretical calculations, and with explaining why the current
matter and vacuum energy densities are so similar (the `cosmic
coincidence' problem).  For these reasons, amongst others, a large
number of alternative cosmological models have been proposed.  These
include models that introduce new energy components to the Universe -
so called `dark energy' models e.g. scalar `quintessence' fields
\citep{Caldwell:98, Zlatev:98, Copeland:98, Steinhardt:99,
Barreiro:00}, K-essence \citep{Chiba:00, Armendariz-Picon:00,
Armendariz-Picon:00b}, tachyon fields \citep{Bagla:02, Copeland:04} and
Chaplygin gas models \citep{Kamenshchik:01, Bento:02}. Other
possibilities include modified gravity theories, motivated by e.g. the
existence of extra dimensions \citep{Dvali:00, Deffayet:02,
Deffayet:02b, Maartens:06, Guo:06} or other modifications of General
Relativity \citep{Capozziello:03, Carroll:03b, Vollick:03, Carroll:04,
Mena:05, Navarro:05, Nojiri:06}, which can also lead to late-time cosmic
acceleration.  The simplicity of the concordance $\Lambda$CDM model,
however, makes it highly attractive. A central goal of modern
observational cosmology is to test whether this model continues to
provide an adequate description of rapidly improving data.

Most current analyses of cosmological data assume General Relativity
and employ the mean matter density of the Universe, $\Omega_{\rm m}$,
and the dark energy equation of state $w$ as model parameters.  Such
analyses are often referred to as `dynamical studies', employing as
they do the Friedmann equations. Other dynamical analyses employ
modified Friedmann equations for a particular gravity model. However,
a purely kinematical approach is also possible that does not assume
any particular gravity theory. Kinematical models provide important,
complementary information when seeking to understand the origin of the
observed late-time accelerated expansion. 

In a pioneering study, \cite{Riess04} measured a transition from a
decelerating to accelerating phase using a simple linear
parameterization of the deceleration parameter $q(z)$, where $q(z)$ is
the dimensionless second derivative of the scale factor, $a(t)$, with
respect to cosmic time. Recently, \cite{Shapiro:05}, \cite{Gong:06}
and \cite{Elgaroy:06} have employed a variety of other
parameterizations, constructed in terms of $q(z)$, to study this
transition. However, since the underlying physics of the transition
are unknown, the choice of a particular parameterization for $q(z)$ is
quite arbitrary.  \cite{Shapiro:05} applied a principal component
analysis of $q(z)$ to the supernovae data of \cite{Riess04} and found
strong evidence for recent, changing acceleration but weak evidence
for a decelerated phase in the past (i.e. weak evidence for a
transition between the two phases). \cite{Elgaroy:06} employed a
Bayesian analysis to the \cite{Riess04} data and the more recent SNLS
supernovae sample of \cite{Astier06}, obtaining a similar result.

In this paper we develop an improved method for studying the
kinematical history of the Universe. Instead of using
parameterizations constructed in terms of $q(z)$, we follow
\cite{Blandford:04} and introduce the cosmic jerk, $j(a)$, the
dimensionless third derivative of the scale factor with respect to
cosmic time. (Here $a$ is the cosmic scale factor, with $a=1/1+z$.)
The use of the cosmic jerk formalism provides a more natural parameter
space for kinematical studies. Our results are presented in terms of
current deceleration parameter $q_{0}$ and $j(a)$, where the latter
can be either constant or evolving. We apply our method to the three
best current kinematical data sets: the`gold' sample of type Ia
supernovae (hereafter SNIa) measurements of \cite{Riess04}, the SNIa
data from the first year of the Supernova Legacy Survey (SNLS) project
\citep{Astier06}, and the X-ray galaxy cluster distance measurements
of \cite{Allen06}. This latter data set is derived from measurements
of the baryonic mass fraction in the largest relaxed galaxy clusters,
which is assumed to be a standard quantity for such systems \citep[see
e.g.][for discussion]{Allen:04}.

In General Relativity $j(a)$ depends in a non-trivial way on both
$\Omega_{\rm m}$ and $w(a)$ \citep{Blandford:04}. In general, there is
no one-to-one mapping between models with constant $j$ and models with
constant $w$. A powerful feature of the standard dynamical approach is
that all $\Lambda$CDM models have $w=-1$ which make it easy to search
for departures from $\Lambda$CDM.  Likewise, the use of the jerk
formalism imbues the kinematical analysis with a similar important
feature in that all $\Lambda$CDM models are represented by a single
value of $j=1$. The use of the jerk formalism thus enables us to
constrain and explore departures from $\Lambda$CDM in the kinematical
framework in an equally effective manner. Moreover, by employing both
the dynamical and kinematical approaches to the analysis of a single
data set, we explore a wider set of questions than with a single
approach. We note that \cite{Sahni:02} and \cite{Alam:03} also drew
attention to the importance of the jerk parameter for discriminating
models of dark energy and/or modified gravity. \cite{Chiba:98} and
\cite{Caldwell:04} also showed its relevance for probing the spatial
curvature of the Universe.

Using the three kinematical data sets mentioned above, we find clear
evidence for a negative value of $q_{\rm 0}$ (current acceleration)
and a positive cosmic jerk, assuming $j$ constant. The concordance
$\Lambda$CDM model provides a reasonable description of the data,
using both the new kinematical and standard dynamical approaches.  We
also search for more complicated deviations from $\Lambda$CDM,
allowing $j(a)$ to evolve as the Universe expands, in an analogous
manner to dynamical studies which allow from time-variation of the
dark energy equation of state $w(a)$. Our analysis employs a Chebyshev
polynomial expansion and a Markov Chain Monte Carlo approach to
explore parameter spaces.  We find no evidence for a time-varying
jerk.

This paper is structured as follows: in section~\ref{dek} we describe
our new kinematical approach. In section~\ref{evoljerk} we describe the
scheme adopted for polynomial expansions of $j(a)$.
Section~\ref{analysis} includes details of the data analysis. 
The results from the application of our method to the supernovae
and X-ray cluster data are
presented in section~\ref{constraints}. Finally, our main 
conclusions are summarized
in section~\ref{discussion}. Throughout this paper, we assume that the
Universe is geometrically flat.

\section{The kinematical and dynamical frameworks for 
late time cosmic acceleration}
\label{dek}

\subsection{Previous work}
\label{previous}

The expansion rate of the Universe can be written in terms of the
Hubble parameter, $H=\dot{a}/a$, where $a$ is the scale factor and
$\dot{a}$ is its first derivative with respect to time.  The current
value of the Hubble parameter is the Hubble Constant, usually written
as $H_{\rm 0}$. Under the action of gravity, and for negligible vacuum
energy, the expansion of the Universe is expected to decelerate at
late times. Contrary to this expectation, in the late 1990s, type Ia
supernovae experiments \citep{Riess:98,Perlmutter:98} provided the
first direct evidence for a late time accelerated expansion of the
Universe. In particular, the present value of the deceleration
parameter, $q_{\rm 0}$, measured from the supernova data was found to
be negative. In detail, the deceleration parameter $q$ is defined as
the dimensionless second derivative of the scale factor

\begin{equation}
q(t)=-\frac{1}{H^2}\left(\frac{\ddot{a}}{a}\right)\,,
\label{dec}
\end{equation}

\noindent and in terms of the scale factor, 

\begin{equation}
q(a)=-\frac{1}{H}(a H)'
\label{deca}
\end{equation}

\noindent where the `dots' and `primes' denote derivatives with
respect to cosmic time and scale factor, respectively.

The current `concordance' cosmological model, $\Lambda$CDM, has been
successful in explaining the SNIa results and all other precision
cosmology measurements to date. Together with it's theoretical
simplicity, this makes the $\Lambda$CDM model very attractive.
However, as discussed in the introduction, the concordance model does
face significant theoretical challenges and a wide-range of other
possible models also provide adequate descriptions of the current data
\citep[see][for an extensive review]{Copeland:06}.

An excellent way to distinguish between models 
is to obtain precise measurements of the
time evolution of the expansion of the Universe. 
Given such data, a number of different analysis
approaches are possible. 
In searching for time evolution in the deceleration parameter, as
measured by current SNIa data, 
\cite{Riess04} assumed a linear
parameterization of $q(z)$,

\begin{equation}
q(z)=q_{\rm 0}+\frac{dq}{dz}z.
\label{riess_dec}
\end{equation}

\noindent These authors measured a change in sign of the deceleration
parameter, from postive to negative approaching the present day, at a
redshift $z_{\rm t}= 0.46\pm0.13$. Using this parameterization for
$q(z)$, the definition of the deceleration parameter given by equation
(\ref{dec}), and integrating over the redshift, we obtain that for
this model the evolution of the Hubble parameter is given in the form

\begin{equation}
E(z)= H(z)/H_{\rm 0} = (1+z)^{(1+q_{\rm 0}-q')}e^{q'z},
\label{soludec}
\end{equation}

\noindent where $q'=dq/dz$.

However, since the origin of cosmic acceleration is unknown, it is
important to recognize that the choice of any particular parameterized
expansion for $q(z)$ is essentially arbitrary.  Indeed, when (or if) a
transition between decelerated and accelerated phases if inferred to
occur can depend on the parameterization used. \cite{Elgaroy:06}
showed that using the linear parameterization described by equation
(\ref{riess_dec}) and fitting to the SNIa data set of \cite{Astier06}
a transition redshift $z_{\rm t}\sim 2.0$ is obtained which,
uncomfortably, lies beyond the range of the data used. 

Transitions between phases of different cosmic acceleration are more
naturally described by models incorporating a cosmic `jerk'. The jerk
parameter, $j(a)$, is defined as the dimensionless third
derivative of the scale factor with respect to cosmic time \citep{Blandford:04}

\begin{equation}
j(t)=\frac{1}{H^3}\left(\frac{\dot{\ddot{a}}}{a}\right)\,,
\end{equation}

\noindent and in terms of the scale factor

\begin{equation}
j(a)=\frac{(a^2 H^2)'' }{2H^2}
\label{defj}
\end{equation}

\noindent where again the `dots' and `primes' denote derivatives with
respect to cosmic time and scale factor, respectively.

In such models, a transition from a decelerating phase at
early times to an accelerating phase at late times occurs for all
models with $q_{\rm 0}<0$ and a positive cosmic jerk. Note that a
Taylor expansion of the Hubble parameter around small redshifts
\citep{Visser:03, Riess04} contains the present value of both the
deceleration and jerk parameters, $q_{\rm 0}$ and $j_{\rm 0}$. Such
Taylor expansions are inappropriate for fitting high-redshift objects
\citep{Blandford:04, Linder:06}, such as those included in the data
sets used here.

\cite{Blandford:04} describe how the jerk parameterization provides a
convenient, alternative method to describe models close to
$\Lambda$CDM. In this parameterization, flat $\Lambda$CDM models have
a constant jerk with $j(a)=1$ (note that this neglects the effects of
radiation over the redshift range of interest, which is also usually
the case when modeling within the dynamic framework).  Thus, any
deviation from $j=1$ measures a departure from $\Lambda$CDM, just as
deviations from $w=-1$ do in more standard dynamical analyses.
Importantly, in comparison to dynamical approaches, however, the
kinematical approach presented here both explores a different set of
models and imposes fewer assumptions. The dynamical approach has other
strengths, however, and can be applied to a wider range of data (e.g.
CMB and growth of structure studies), making the kinematical and
dynamical approaches highly complementary.

It is interesting to note that, in principle, any particular dynamical
parameter space will have its own physical limits. For instance,
within the dynamical $(\Omega_{\rm m},w)$ plane, models with $w<-1$,
known as `phantom' dark energy models, violate the dominant energy
condition \citep{Carroll:03, Onemli:04} and present serious problems
relating to the treatment of dark energy perturbations \citep{Hu:04,
  Vikman:05, Caldwell:05, Zhao:05} when $w(z)$ crosses the boundary
$w=-1$. Current data allow models with $w<-1$ \citep{Weller:03,
  Riess04, Allen:04, Astier06, Spergel:06, Cabre:06} and models in
which $w(z)$ crosses the boundary $w=-1$ \citep{Jassal:04,
  Corasaniti:04a, Seljak:04, Rapetti:05, Upadhye:05, Zhao:06}.
However, another dynamical parameter space, coming e.g. from a
different gravity theory, might not pathologically suffer from such
boundaries around the models allowed for current data.

Since the $(q_{\rm 0},j)$ plane (see below) is purely kinematical,
i.e. no particular gravity theory is assumed, we are not forced to
interpret $j=1$, or any locus in this plane, as a barrier.  Note,
however, that caution is required in extending the results from the
kinematical analysis beyond the range of the observed data
\citep[see][for details]{Blandford:06}. For example, inappropriately
extending a jerk model to very high redshifts could imply an
unphysical Hubble parameter at early times, i.e., these models do not
have a Big Bang in the past.
 
\subsection{A new kinematical framework}
\label{kframe}

For our kinematical analysis, we first calculate $H(a)$ given
$j(a;\mathcal{C})$ where $\mathcal{C}=(c_{\rm 0}, c_{\rm 1}, ...,
c_{\rm N})$ is the selected vector of parameters used to describe the
evolution of $j(a)$ (see below). Following \cite{Blandford:04} we
rewrite the defining equation for the jerk parameter (\ref{defj}) in a
more convenient form

\begin{equation}
a^2V''(a)-2j(a)V(a)=0
\label{jerk_dif_eq}
\end{equation}

\noindent where $'$ denotes derivative with respect to $a$
and $V(a)$ is defined as

\begin{equation}
V(a)=-\frac{a^2H^2}{2H^2_{\rm 0}}\,.
\label{defi}
\end{equation}

\noindent We specify the two constants of integration required by
(\ref{jerk_dif_eq}) in terms of the present Hubble parameter $H_0$ and
the present deceleration parameter $q_0$ as follows

\begin{equation}
V(1)=-\frac{1}{2}\,,
\qquad V'(1)=q_0\,,
\label{initial}
\end{equation}

\noindent where $a(t_0) =1$ at the present time $t_0$. Here the first
condition comes from $H(1)=H_{\rm 0}$ and the second from

\begin{equation}
V'(1)=-\frac{H'_{\rm 0}}{H_{\rm 0}}-1=q_{\rm 0}\,.
\label{c2}
\end{equation}

\noindent The Hubble parameter, $H(a)$, obtained from equations
(\ref{jerk_dif_eq}), (\ref{defi}) and (\ref{initial}) is used to
calculate the angular diameter ($d_{\rm A}$) and luminosity ($d_{\rm
  L}$) distances for a flat Friedmann-Robertson-Walker-Lema\^{i}tre (FRWL)
metric

\begin{equation}
d_{\rm A}(a)= a^2\,d_{\rm L}(a)= \frac{c}{H_{\rm 0}}\,a\,\int_{a}^{1}\frac{1}{a^2 E(a)}\,da,
\label{distances}
\end{equation}

\noindent where $c$ is the speed of light. These theoretical
distances $d_{\rm L}(a)$ and $d_{\rm A}(a)$ are then used in the data
analysis (see section~\ref{analysis}).

Our framework provides a simple and intuitive approach for kinematical
studies. For models with $q_0<0$ ($>0$), the Universe is currently
accelerating (decelerating). Models with $q_0<0$ and $j(a)=1$
(constant) are currently accelerating and have the expansion evolving
in a manner consistent with $\Lambda$CDM. Any significant departure
from $j=1$ indicates that some other mechanism is responsible for the
acceleration. 

\subsection{Standard dynamical framework}

For comparison purposes, we have also carried out a standard dynamical
analysis of the data in which we employ a dark energy model with a
constant dark energy equation of state, $w$.  From energy conservation
of the dark energy fluid and the Friedmann equation, we obtain the
evolution of the Hubble parameter, $H(z)=H_{\rm 0}\,E(z)$,

\begin{equation}
  E(z) = [ \Omega_{\rm m} (1+z)^3 + (1-\Omega_{\rm m}) (1+z)^{3(1+w)}]^{1/2}, 
\label{whubble}
\end{equation}

\noindent where $\Omega_{\rm m}$ is the mean matter density in units
of the critical density. As with the kinematical analysis, we assume
flatness and neglect the effects of radiation density. In this
framework, models with a cosmological constant have $w=-1$ at all
times.

\section{Evolving jerk models}
\label{evoljerk}

Our analysis allows for the possibility the cosmic jerk
parameter, $j(a)$ may evolve with the scale factor. We have restricted our
analysis to the range of $a$ where we have data, $[a_{\rm min}=0.36,
a_{\rm max}=1]$. In searching for
possible evolution, our approach 
is to adopt $\Lambda$CDM as a base model and
search for progressively more complicated deviations from this. 
We begin by allowing a constant
deviation $\Delta j$ from $\Lambda$CDM ($j=1$). 
For this model, it is possible to solve the jerk
equation (\ref{jerk_dif_eq}) analytically.  Using the initial
conditions listed in (\ref{initial}), we obtain

\begin{equation}
  V(a)=-\frac{\sqrt{a}}{2}\, \left[\left(\frac{p-u}{2p}\right) a^{p} + \left(\frac{p+u}{2p}\right) a^{-p}\right]
\label{const_jerk}
\end{equation}

\noindent where $p\equiv (1/2)\sqrt{(1+8j)}$ and $u\equiv 2(q_{\rm
  0}+1/4)$. Note that in the ($q_{\rm 0}$,$j$) plane for

\begin{equation}
  j< \left\{ \begin{array}{ll} 
      q_{\rm 0}+2q_{\rm 0}^2 & q_{\rm 0}<-1/4\\
      -1/8 & q_{\rm 0}>-1/4
    \end{array} \right\}
\label{curvebigbang}
\end{equation}

\noindent there is no Big Bang in the past~\footnote{Allowed ($q_{\rm
    0}$,$j$) values are those for which the equation $V(a)=0$ has no
  solution in the past ($a<1$) \citep{Blandford:06}.}. The models
allowed by our combined data sets do not cross this boundary.

For the next most complicated possible deviation from $\Lambda$CDM, we
have $j(a;\mathcal{C})=j^{\rm \Lambda CDM}+\Delta j(a;\mathcal{C})$.
Here $j^{\rm \Lambda CDM}=1$ and $j(a;\mathcal{C})$ is the cosmic jerk
for the cosmology in question. In order to meaningfully increase the
number of parameters in the vector $\mathcal{C}$, we employ a
framework constructed from Chebyshev polynomials.  The Chebyshev
polynomials form a basis set of polynomials that can be used to
approximate a given function over the interval $[-1,1]$. We rescale
this interval to locate our function $\Delta j(a;\mathcal{C})$ in the
range of scale factor where we have data:
 
\begin{equation}
  a_c\equiv \frac{a-(1/2)(a_{\rm min}+a_{\rm max})}{(1/2)(a_{\rm max}-a_{\rm min})}\, ,
\label{rescale}
\end{equation}

\noindent where $a$ is the scale factor in the range of interest and
$a_{\rm c}$ is Chebyshev variable. The trigonometric expression for a
Chebyshev polynomial of degree $n$ is given by 

\begin{equation}
  T_{\rm n}(a_{\rm c})=\cos({n \arccos a_{\rm c}}).
\label{tricheb}
\end{equation}

\noindent These polynomials can also be calculated using the
recurrent formula

\begin{equation}
 T_{\rm n+1}(a_{\rm c}) = 2a_{\rm c}T_{\rm n}(a_{\rm c})-T_{\rm n-1}(a_{\rm c}), \qquad n\geq 1 \, ,
\label{recurrent}
\end{equation}

\noindent where $T_{\rm 0}(a_{\rm c})=1$ and, for example, the next three
orders are $T_{\rm 1}(a_{\rm c})=a_{\rm c}$, $T_{\rm 2}(a_{\rm c})=2{a_{\rm c}}^2-1$, $T_{\rm
  3}(a_{\rm c})=4{a_{\rm c}}^3-3a_{\rm c}$, etc. Using a weighted combination of these
components, any arbitrary function can be approximately reconstructed.
The underlying deviation from $\Lambda$CDM can be expressed as
 
\begin{equation}
  \Delta j(a;\mathcal{C})\simeq \sum^{\rm N}_{\rm n=0}c_{\rm n}T_{\rm n}(a_{\rm c}) 
  \label{aproxfunct}
\end{equation}

\noindent where the weighting coefficients form our vector of
parameters, $\mathcal{C}=(c_{\rm 0}, c_{\rm 1}, ..., c_{\rm N})$.
Thus, using equation (\ref{aproxfunct}) we produce different
parameterizations for increasing $N$. With higher $N$'s we allow a
more precise exploration of the $[q_{\rm 0},j(a;\mathcal{C})]$
parameter space. However, it is clear that this process will be
limited by the ability of the current data to distinguish between such
models. In order to judge how many orders of polynomials to include,
we quantify the improvements to the fits obtained from the inclusion
of progressively higher orders in a variety of ways (see below). In
general, we find that models with a degree of complexity beyond a
constant jerk are not required by current data.

We note that approches other than expanding $\Delta j$ in Chebyshev
polynomials are possible, e.g. one could include the dimensionless
fourth derivative of the scale factor as a model parameter. However,
since $\Lambda$CDM does not make any special prediction for the value
of this derivative, we prefer to use our general expansion in $\Delta
j$ here.


\section{Data and analysis methods}
\label{analysis}

\subsection{Type Ia supernovae data}

For the analysis of SNIa data, we use both the `gold' sample of
\cite{Riess04} and the first year SNLS sample of \cite{Astier06}. The
former data set contains 157 \footnote{\cite{Riess04} presented 16 new
  Hubble Space Telescope (HST) SNIa, combined with 170 previously
  reported SNIa from ground-based data.  They identified a widely used
  ``high-confidence'' subset, usually referred to as the {\it gold
    sample}, which includes 14 HST SNIa.} SNIa, where a subset of 37
low-redshift objects are in common with the data of \cite{Astier06}.
\cite{Astier06} contains 115 \footnote{71 SNLS objects, plus 44
  previously reported nearby objects.} objects.  We use the
measurements of \cite{Astier06} for objects in common between the
studies.  Thus, combining both data sets we have 120 SNIa from the
\cite{Riess04} gold sample (157 minus the 37 low-redshift objects in
common) and 115 SNIa from \cite{Astier06}.

The two SNIa studies use different light-curve fitting methods. In
order to compare and combine the data, we fit the observed distance
moduli $\mu^{\rm obs}(z_{\rm i}) = m^{\rm obs}(z_{\rm i})-M$, where
$m$ is the apparent magnitude at maximum light after applying galactic
extinction, K-correction and light curve width-luminosity corrections,
and $M$ is the absolute magnitude, with the theoretical predictions,
$\mu^{\rm th}(z_{\rm i}) = m^{\rm th}(z_{\rm i})-M = 5\log_{\rm 10}
D_{\rm L}(z_{\rm i};\theta)+\mu_{\rm 0}$, where $D_{\rm L}=H_{\rm
  0}\,d_{\rm L}$ is the $H_{\rm 0}$-free luminosity distance,
$\mu_{\rm 0} = 25 - 5log_{\rm 10}H_{\rm 0}$ and $m_{\rm 0}\equiv
M+\mu_{\rm 0}$ is a ``nuisance parameter'' which contains both the
absolute magnitude and $H_{\rm 0}$.

For the $[q_{\rm 0},j(a;\mathcal{C})]$ parameter space, the luminosity
distance $d_{\rm L}(z;\theta)$ is directly obtained integrating the
solution of the differential equation (\ref{jerk_dif_eq}) with the
definition (\ref{distances}) as presented in subsection~\ref{kframe}.
For models using linear parameterization of $q(z)$ and/or dynamical
models with $\Omega_{\rm m}$ and $w$, we plug the equations
(\ref{soludec}) and (\ref{whubble}), respectively, into the equation
describing the luminosity distance for a flat FRWL metric, in units of
megaparsecs

\begin{equation}
d_{\rm L}(z;\theta)= \frac{c (1+z)}{H_{\rm 0}}\int_0^z\frac{dz}{E(z;\theta)}
\label{ludist}
\end{equation}

\noindent where the speed of light, c, is in km s$^{-1}$ and the present
Hubble parameter, $H_{\rm 0}$, in  km(s Mpc)$^{-1}$. Here the vectors of
parameters for each model are $\theta=(q_{\rm 0},dq/dz)$ and
$\theta=(\Omega_{\rm m},w)$ respectively. For the gold sample data of
\cite{Riess04}, we use the extinction-corrected distance moduli,
$\mu^{\rm obs}(z_{\rm i})$ and associated errors, $\sigma^2_{\rm i}$.
For the SNLS data of \cite{Astier06} we use the rest-frame-B-band
magnitude at maximum light $m^{\rm *}_{\rm B}(z_{\rm i})$, the stretch
factor $s_{\rm i}$ and the rest-frame color $c_{\rm i}$ to obtain
$\mu^{\rm obs}(z_{\rm i})= m^{*}_{\rm B}(z_{\rm i}) - M + \alpha
(s_{\rm i}-1)- \beta c_{\rm i}$. These values were derived from the
light curves by \cite{Astier06}, who also provide best-fitting values
for $\alpha=1.52 \pm 0.14$ and $\beta=1.57 \pm 0.15$.

For both SNIa data sets, we have

\begin{equation}
  \chi^2(\theta;m_{\rm 0})=\sum_{\rm SNIa}\frac{[\mu^{\rm th}
    (z_{\rm i};\theta,\mu_{\rm 0})-\mu^{\rm obs}(z_{\rm i};\theta,M)]^2}{\sigma^2_{\rm i}},
\label{chisquare}
\end{equation}

\noindent where the dispersion associated with each data point,
$\sigma^2_{\rm i}=\sigma^2_{\mu_{\rm i, obs}}+\sigma^2_{\rm int,
  i}+\sigma^2_{\rm v, i}$. Here $\sigma^2_{\mu_{\rm i, obs}}$ accounts
for flux uncertainties, $\sigma^2_{\rm int, i}$ accounts for
intrinsic, systematic dispersion in SNIa absolute magnitudes and
$\sigma^2_{\rm v, i}$ accounts for systematic scatter due to peculiar
velocities. The SNLS analysis includes an intrinsic dispersion of
$0.13104$
magnitudes\footnote{http://snls.in2p3.fr/conf/papers/cosmo1/} and a
peculiar velocity scatter of $300$ km/s.  The gold sample analysis
includes $400$ km/s peculiar velocity scatter, with an additional
$2500$ km/s added in quadrature for high redshift SNIa.

We marginalise analytically over $m_{\rm 0}$

\begin{equation}
  \tilde{\chi}^{2}(\theta)=
  -2\ln\int^{\infty}_{-\infty}{\exp\left(-\frac{1}{2}\chi^{2}
      (\theta,m_{\rm 0})\right)}dm_{\rm 0}
\end{equation}

\noindent obtaining

\begin{equation}
\tilde{\chi}^2=\ln\left(\frac{c}{2\pi}\right)+a-\frac{b^{2}}{c},
\label{margin}
\end{equation}

\noindent where

\begin{equation}
a=\sum_{\rm SNIa}\frac{[5\log_{\rm 10} D_{\rm L}(z_{\rm i};
\theta)-m^{\rm obs}(z_{\rm i})]^{2}}{\sigma_{\rm i}^{2}},
\end{equation}

\begin{equation}
  b=\sum_{\rm SNIa}\frac{5\log_{\rm 10} D_{\rm L}(z_{\rm i};
    \theta)-m^{\rm obs}(z_{\rm i})}{\sigma_{\rm i}^{2}}, 
    \qquad c=\sum_{\rm SNIa}\frac{1}{\sigma_{\rm i}^{2}}.
\end{equation}

\noindent Note that the absolute value of $\chi^2=a-(b^2/c)$. For the
analysis in the standard dynamic framework, our results agree with
those of \cite{Riess04} and \cite{Astier06}, and the comparison work
of \cite{Nesseris:05}.

\subsection{X-ray cluster data}

For the analysis of cluster X-ray gas mass fractions, we use the data of
\cite{Allen06}, which contains 41 X-ray luminous, relaxed galaxy
clusters, including 26 previously studied \cite{Allen:04}. [Some
of the original 26 have since been revisited by the Chandra X-ray
observatory leading to improved constraints \cite[for details
see][]{Allen06}.] The new X-ray data set spans a redshift interval
$0.06<z<1.07$. Our analysis follows the method of \cite{Allen:04},
fitting the apparent redshift evolution of the cluster gas fraction
with the expression

\begin{equation}
  f_{\rm gas}^{\rm ref}(z_{\rm i}) = \mathcal{F}\, R^{\rm ref}(z_{\rm i}),\quad
  R^{\rm ref}(z_{\rm i})\equiv\left[\frac{d_{\rm A}^{\rm ref}(z_{\rm i})}{D_{\rm A}(z_{\rm i})}\right]^{1.5},
\label{eq:fgas}
\end{equation}

\noindent where $\mathcal{F}= (b\, \Omega_{\rm b}H_{\rm
  0}^{1.5})/[(1+0.19\sqrt{h})\, \Omega_{\rm m}]$ is the normalization
of the $f_{\rm gas}(z)$ curve, $d_{\rm A}$ and $d_{\rm A}^{\rm
  ref}(z)$ are the angular diameter distances ($d_A=d_L/(1+z)^2$) to
the clusters for a given cosmology and for the reference $\Lambda$CDM
cosmology (with $H_{\rm 0}=70$ km(s Mpc)$^{-1}$ and $\Omega_{\rm m}=0.3$)
respectively, and $D_{\rm A}=H_{\rm 0}\,d_{\rm A}$ is the $H_{\rm
  0}$-free angular diameter distance. For the kinematical approach we
treat the normalization $\mathcal{F}$ as a single `nuisance'
parameter, which we marginalize over in the MCMC chains.

For the $dynamical$ analysis of the same X-ray data, we follow
\cite{Allen:04} and employ Gaussian priors on the present value of the
Hubble parameter $H_{\rm 0}=72\pm 8$ km(s Mpc)$^{-1}$
\citep{Freedman:01}, the mean baryon density $\Omega_{\rm
  b}h^2=0.0214\pm0.0020$ \citep{Kirkman:03} and the X-ray bias factor
$b=0.824\pm0.089$ [determined from the hydrodynamical simulations of
\cite{Eke:97}, including a 10 per cent allowance for systematic
uncertainties]. The application of these priors leads to an additional
constraint on $\Omega_{\rm m}$ from the normalization of the $f_{\rm
  gas}(z)$ curve. Since the kinematical approach does not constrain
$\Omega_{\rm m}$, the kinematical analysis does not involve these
priors and draws information only from the shape of the $f_{\rm
  gas}(z)$ curve. The dynamical analysis, in constrast, extracts
information from both the shape $and$ normalization.

\begin{table*}
\begin{center}
  \caption{The marginalised median values and 68.3 per cent confidence
    intervals obtained analysing each data set and all three data sets
    together. We show these results for two kinematical models: using
    only $q_{\rm 0}$ ($\mathcal{Q}$ model) and extending this
    parameter space with the jerk parameter $j$ ( $\mathcal{J}$
    model). We quote $\chi^{2}$ per degree of freedom for each model
    and three different statistical tests to quantify the significance
    of extending the parameter space from $\mathcal{Q}$ ($q_{\rm 0}$)
    to $\mathcal{J}$ ($q_{\rm 0}$,$j$). We quote the difference in
    $\Delta \chi^2_{\rm \mathcal{JQ}}$, the probability given by
    F-test, the difference in the Bayesian Information Criterion (BIC)
    and $\ln B_{\rm \mathcal{JQ}}$ (where $B_{\rm \mathcal{JQ}}$ is
    the Bayes factor between the two models). Note that combining all
    three data sets we obtain a significant preference for the
    $\mathcal{J}$ model within all three tests.}
\label{tab}
\begin{tabular}{ l c c c c c c c c c c c}
\hline
\multicolumn{1}{c}{} &
\multicolumn{2}{c}{$\mathcal{Q}$ model} &
\multicolumn{1}{c}{} &
\multicolumn{3}{c}{$\mathcal{J}$ model} &
\multicolumn{1}{c}{} &
\multicolumn{4}{c}{Improvement} \\
\hline
Data set & $q_{\rm 0}$ & $\chi^{2}_{\rm \mathcal{Q}}/dof$& & $q_{\rm 0}$ & $j$ & $\chi^{2}_{\rm \mathcal{J}}/dof$& & $\Delta \chi^2_{\rm \mathcal{JQ}}$ & F-test & $\Delta BIC$ & $\ln B_{\rm \mathcal{JQ}}$ \\

\noalign{\vskip 5pt}

Clusters &  $-0.55\pm 0.14$  & 39.6/39 & & $-0.61^{+0.38}_{-0.41}$ & $0.51^{+2.55}_{-2.00}$ & 39.6/38 & & 0.01& $5.6\%$ & -3.7 & -3.2 \\

\noalign{\vskip 5pt}

SNLS SNIa &  $-0.417\pm 0.062$  & 112.1/113 & & $-0.65\pm 0.23$& $1.32^{+1.37}_{-1.21}$ & 111.0/112 & & 1.1 &$69.4\%$ & -3.6 & -2.5 \\

\noalign{\vskip 5pt}

Gold SNIa &  $-0.289\pm 0.062$  & 182.8/155 & & $-0.86\pm 0.21$& $2.75^{+1.22}_{-1.10}$ & 174.6/154 & & 8.2 &$99.1\%$ & 3.1 & 1.2 \\

\noalign{\vskip 5pt}

Gold+SNLS+Cl &  $-0.391\pm 0.045$  & 300.8/272 & & $-0.81\pm 0.14$ & $2.16^{+0.81}_{-0.75}$ & 290.1/271 & & 10.7 &$99.8\%$ & 5.1 & 3.0 \\

\hline
\end{tabular}
\end{center}
\end{table*} 

\begin{figure*}
\includegraphics[width=2.87in]{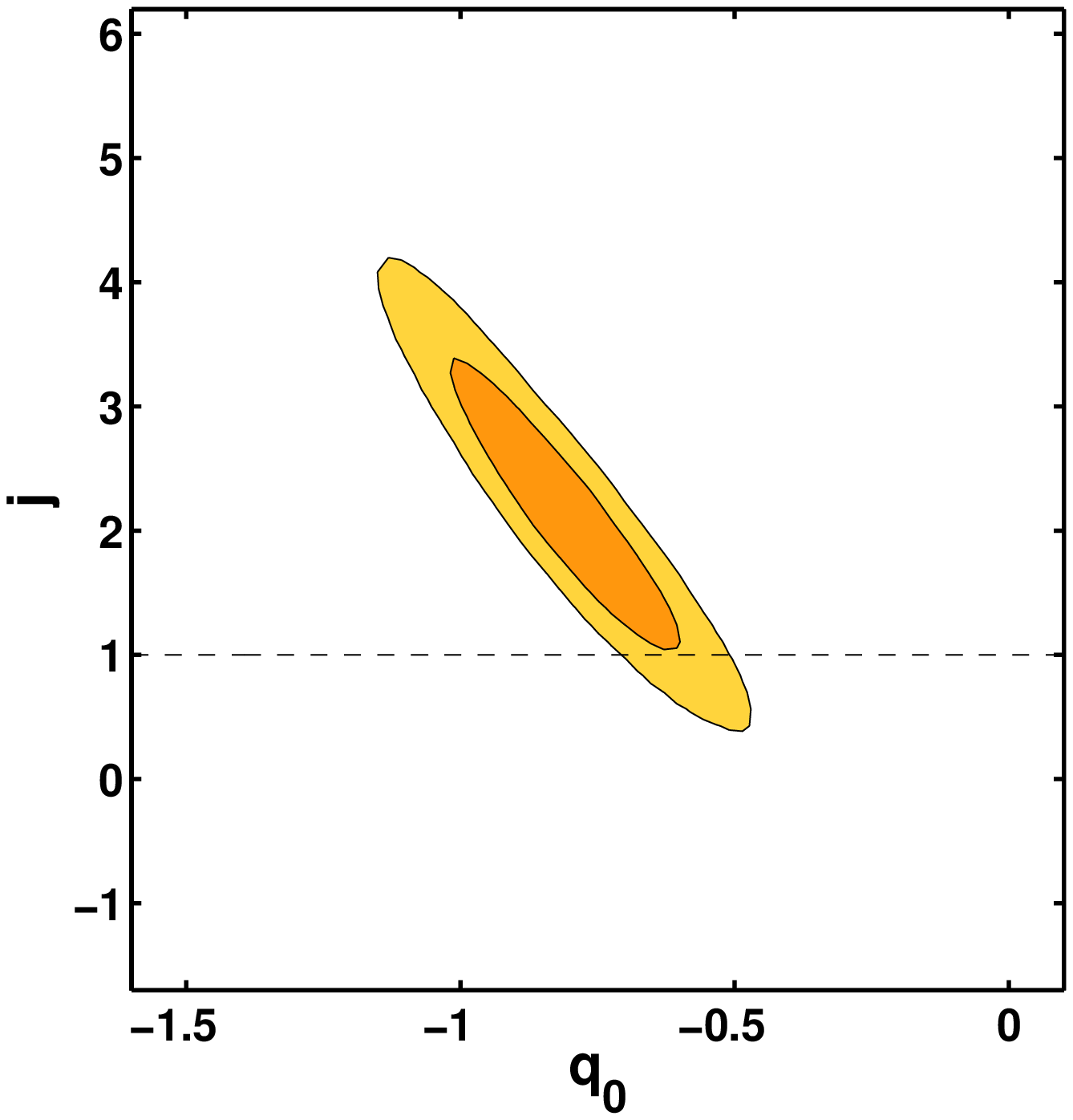}\hspace{0.6cm}
\includegraphics[width=2.9in]{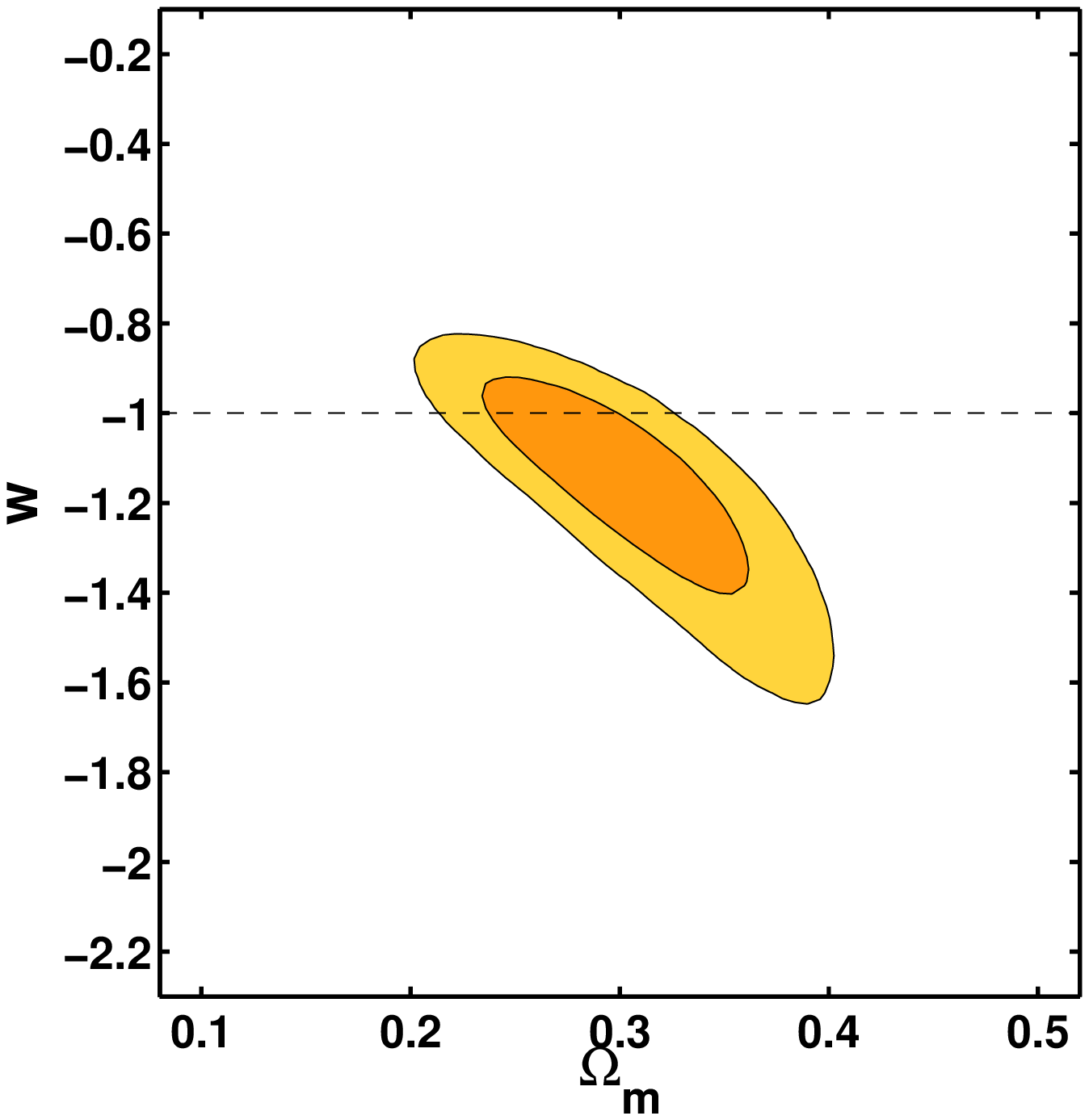}\vspace{0.5cm}
\includegraphics[width=2.87in]{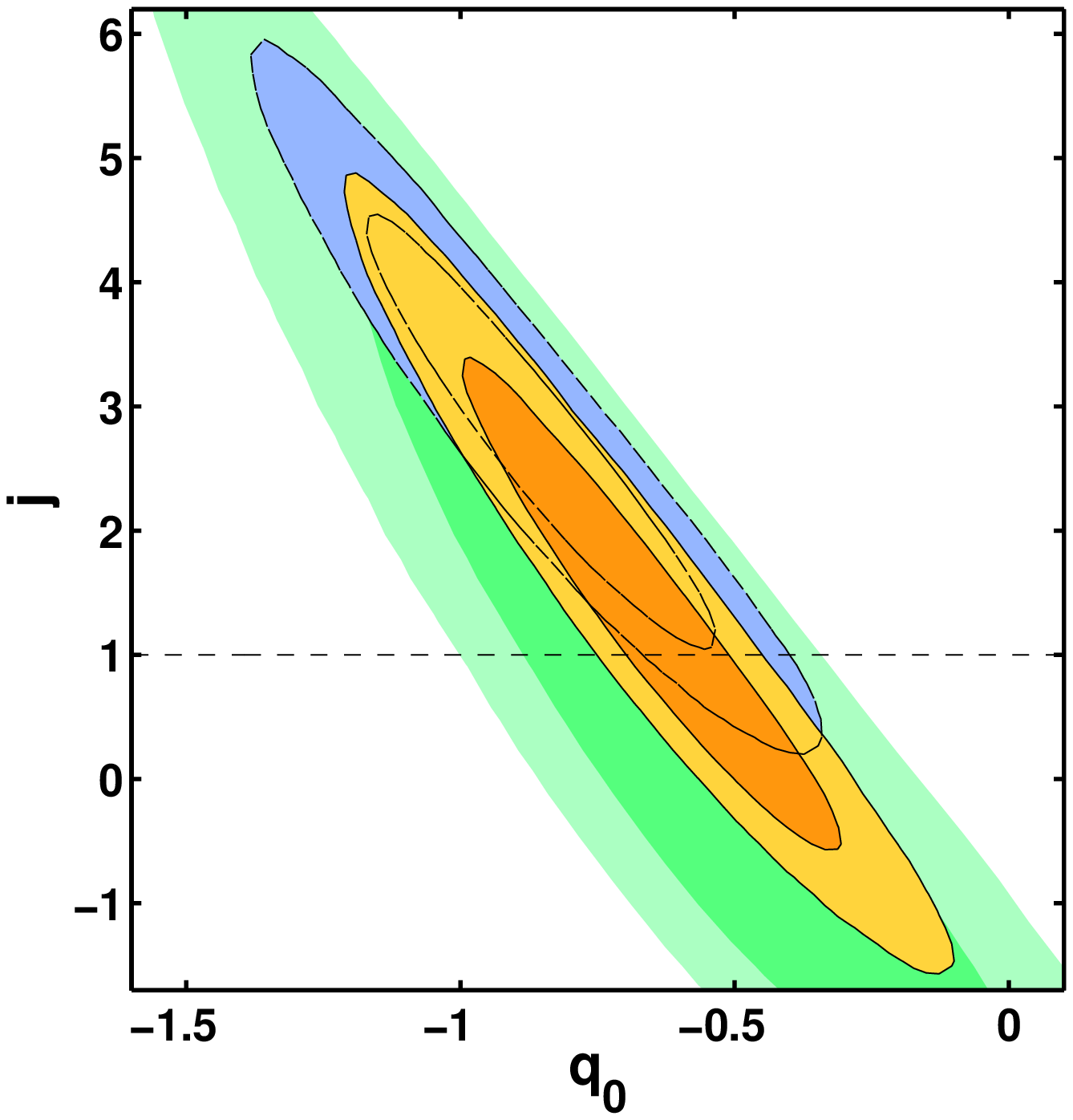}\hspace{0.6cm}
\includegraphics[width=2.9in]{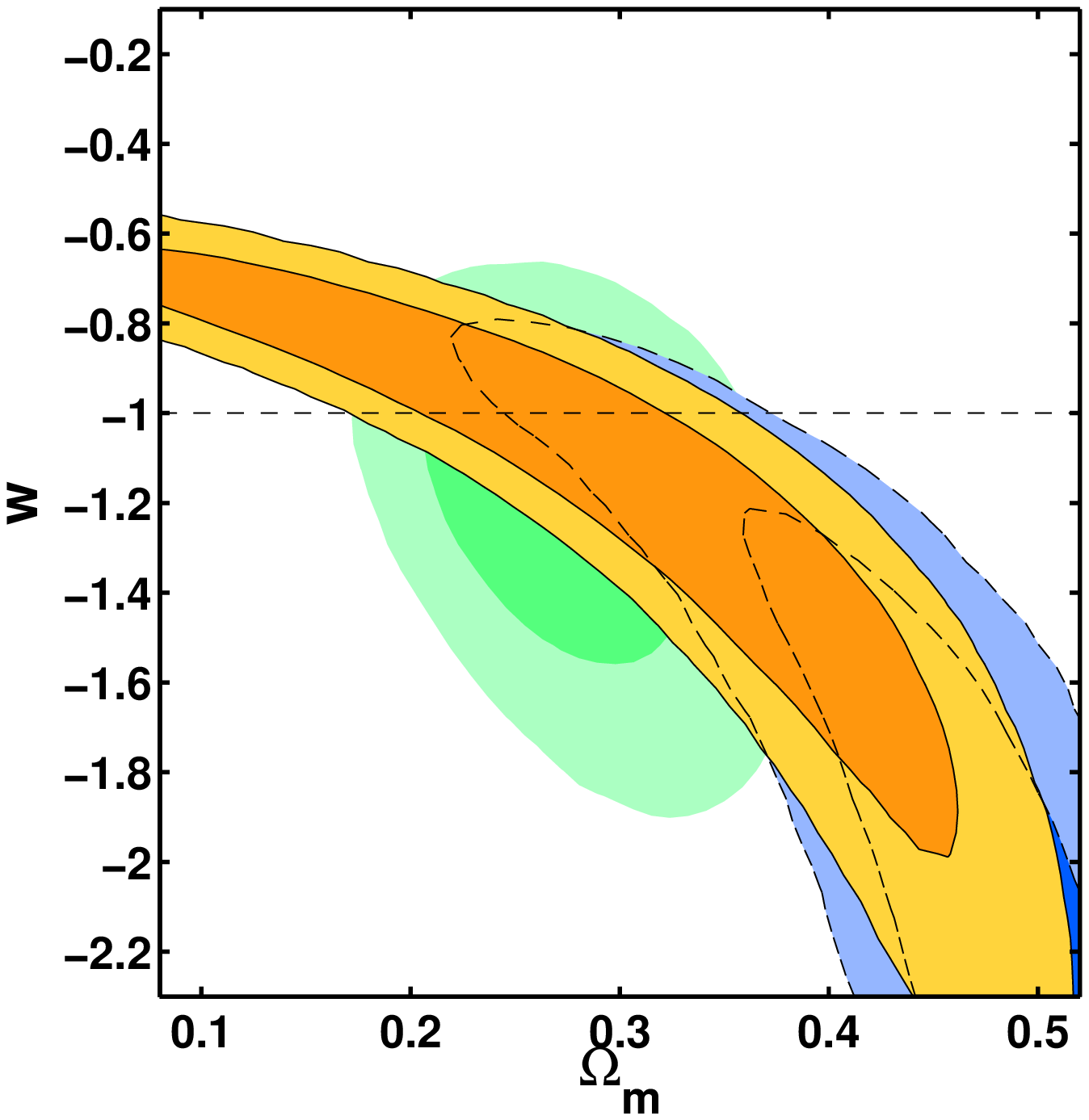}
\caption{A summary of the results from the kinematical (left panels)
  and dynamical (right panels) analyses. The top left panel shows the
  68.3 and 95.4 per cent confidence limits in the ($q_{\rm 0}$,$j$)
  plane for the kinematical model with a constant jerk, $j$, obtained
  using all three data sets: both SNIa data sets
  (Riess et al. 2004; Astier et al. 2005) and the cluster $f_{\rm gas}$ data of
  Allen et al. (2006). The top right panel shows the results in the
  ($\Omega_{\rm m}$,$w$) plane obtained using the same three data sets
  and assuming HST, BBNS and $b$ priors. (Note that the kinematical
  analysis does not use the HST, BBNS and $b$ priors). The dashed
  lines show the expectation for a cosmological constant model in both
  formalisms ($j=1$, $w=-1$, respectively). The bottom panels show the
  confidence contours in the same planes for the individual data sets:
  the SNLS SNIa data (orange contours), the Riess et al. (2004) `gold'
  SNIa sample (blue contours) and the cluster $f_{\rm gas}$ data
  (green contours).  Here, the dashed lines again indicate the
  cosmological constant model.}
\label{dy_ki}
\end{figure*}  

\subsection{Markov Chain Monte Carlo analysis}

For both the kinematical and dynamical analyses,
we sample the posterior probability distributions for all parameter
spaces using a Markov Chain Monte Carlo (MCMC) technique.  This
provides a powerful tool for cosmological studies, allowing the
exploration of large multi-dimensional parameter spaces.  In detail, we use
the Metropolis-Hastings algorithm implemented in the \COSMOMC
~\footnote{http://cosmologist.info/cosmomc/} code of \cite{Lewis:02}
for the dynamic formalism, and a modified version of this code for the
kinematic analysis.

Our analysis uses four MCMC chains for each combination of model and
data. We ensure convergence by applying the Gelman-Rubin criterion
\citep{Gelman:92}, where the convergence is deemed acceptable if the
ratio of the between-chain and mean-chain variances satisfies $R-1 \lt
0.1$. In general, our chains have $R-1\ll 0.1$.

\subsection{Hypothesis testing in the kinematical analysis: 
how many model parameters are required?}
\label{sign}

In the first case, we examined a kinematical model in which the
deceleration parameter $q_{\rm 0}$ was included as the only
interesting free parameter [see equation (\ref{soludec}) with $q'=0$].
This is hereafter referred to as the model $\mathcal{Q}$. As detailed
in section~\ref{dek}, we next introduced the jerk parameter, $j$, as
an additional free parameter, allowing it to take any constant value.
We refer to this as model $\mathcal{J}$, which has the interesting
free parameters, $q_{\rm 0}$ and $j$.  Note that model $\mathcal{J}$
includes the set of possible $\Lambda$CDM models, which all have
constant $j=1$.

We next explored a series of models that allow for progressively more
complicated deviations from $\Lambda$CDM.  In each case, the
improvement obtained with the introduction of additional model
parameters, has been gauged from the MCMC chains using a variety of
statistical tests. In the first case, we follow a frequentist approach
and use the F-test, for which

\begin{equation}
F=\frac{\Delta \chi^2}{\chi^2_{\rm \nu} \, \Delta m}, 
\label{ftest}
\end{equation}

\noindent where $\Delta \chi^2$ is the difference in the minimum
$\chi^2$ between the two models, $\chi^2_{\rm \nu}$ is the reduced
$\chi^2$ ($\chi^2/\nu$, where $\nu$ is number of degrees of freedom of
the fit, $dof$) of the final model, and $\Delta m$ is the difference
in the number of free parameters in the two models. Given $\Delta m$
and $\nu$, we calculate the probability that the new model would give
$\Delta \chi^2 \ge F \, \chi^2_{\rm \nu} \, \Delta m$ by random
chance. This allows us to quantify the significance of the model
extension.

The Bayesian Information Criterion (BIC) provides a more stringent
model selection criterion and is an approximation to the Bayesian
Evidence \citep{Schwarz:78}. The BIC is defined as

\begin{equation}
BIC=-2 \ln \mathcal{L}+k\ln N
\label{bic}
\end{equation}

\noindent where $\mathcal{L}$ corresponds to the maximum likelihood
obtained for a given model (thus, $-2 \ln \mathcal{L}$ is the minimum
$\chi^2$), $k$ is the number of free parameters in the model and $N$
is the number of data points.  Values for $\Delta BIC<2$ between two
models are typically considered to represent weak evidence for an
improvement in the fit. $\Delta BIC$ between 2 and 6 indicates
`positive evidence' for an improvement, and values greater than 6
signify `strong evidence' for the model with the higher BIC
\citep{Jeffreys:61, Kass:95, Mukherjee:98, Liddle:04}.

Finally, we have compared the full posterior probability distributions
for different models, using the Bayes Factor to quantify the
significance of any improvement in the fit obtained. The Bayes Factor
is defined as the ratio between the Bayesian evidences of the two
models \citep{Kass:95}.  If $P(D|\theta,M)$ is the probability of the
data $D$ given a model $M$, the Bayesian evidence is defined as the
integral over the parameter space, $\theta$

\begin{equation}
E(M)\equiv P(D|M)=\int d\theta P(D|\theta,M)\, P(\theta|M),
\label{bev}
\end{equation}

\noindent where $P(\theta|M)$ is the prior on the set parameters
$\theta$, normalised to unity. We employ top hat priors
for all parameters and evaluate the integrals using the
MCMC samples:

\begin{equation}
E(M)\sim \frac{1}{N\Delta \theta}\sum^{\rm N}P(D|\theta_{\rm n},)
\label{samplebe}
\end{equation} 

\noindent where $\Delta \theta$ is the volume in the parameter space
selected to have probability $1$ within the top hat priors, $N$ is the
number of MCMC samples and $\theta_{\rm n}$ the sampled parameter
space. Note that $\sum^{\rm N} P(D|\theta_{\rm n})\,$ is the expected
probability of the data in the posterior distribution
\citep{Lewis:02}.  The evidence of the model $E(M)$ can be estimated
trivially from the MCMC samples as the mean likelihood of the samples
divided by the volume of the prior. It is clear, though, that this
volume will depend on our selection of the top hat priors. In order to
be as objective as possible, within the Bayesian framework, we use the
same priors for parameters in common between the two models involved
in a comparison. For parameters not in common, we calculate their
volumes subtracting their maximum and minimum values in the MCMC
samples.

The Bayes factor between two models $M_{\rm 0}$ and $M_{\rm 1}$ is
$B_{\rm 01}=E(M_{\rm 0})/E(M_{\rm 1})$. If $\ln B_{\rm 01}$ is
positive, $M_{\rm 0}$ is `preferred' over $M_{\rm 1}$. If $\ln B_{\rm
  01}$ is negative, $M_{\rm 1}$ is preferred over $M_{\rm 0}$.
Following the scale of \cite{Jeffreys:61}, if $0<\ln B_{\rm 01}< 1$
only a ``bare mention'' of the preference is considered warranted. If
$1< \ln B_{\rm 01}< 2.5$, the preference is regarded as of
``substantial'' significance. If $2.5< \ln B_{\rm 01}< 5$ the
significance is considered to be going from ``strong'' to ``very
strong''.

\section{RESULTS}
\label{constraints}

\subsection{Comparison of constant jerk and constant $w$ models}
\label{jcontswconst}

\begin{table}
\begin{center}
  \caption{The marginalised median values and 68.3 per cent confidence
    intervals obtained analysing all three data sets together. We show
    the results for the constant $j$ model (kinematical) and the
    constant $w$ model (dynamical) and their corresponding $\chi^{2}$
    per degree of freedom.}
\label{tab:kidy}
\begin{tabular}{ l c c}



\hline



Approach & Model parameters & $\chi^{2}/dof$\\

\hline

\noalign{\vskip 5pt}

Kinematical & $q_{\rm 0}=-0.81\pm 0.14, {\hskip 2pt} j=2.16^{+0.81}_{-0.75}$ & 290.1/271 \\


\noalign{\vskip 5pt}

Dynamical & $\Omega_{\rm m}=0.306^{+0.042}_{-0.040}, {\hskip 2pt} w=-1.15^{+0.14}_{-0.18}$ & 291.7/272 \\

\hline
\end{tabular}
\end{center}
\end{table} 

We first examine the statistical improvement obtained in moving from
the simplest kinematical model $\mathcal{Q}=[q_{\rm 0}]$, in which
$q_0$ is the only interesting free parameter, to model
$\mathcal{J}=[q_{\rm 0},j(c_{\rm 0})]$, where we include constant jerk
$j=1+c_{\rm 0}$ (i.e. we allow $j$ to take values other than zero).
The results obtained, using the three statistical tests described in
subsection~\ref{sign} applied to each data set alone and for all three
data sets together are summarized in Table~\ref{tab}. We find that the
`gold' sample is the only data set that, on it's own, indicates a
`substantial' preference for model $\mathcal{J}$ over model
$\mathcal{Q}$ according to the Bayes factor test. Note that this is
not only due to the fact that the `gold' sample extends to higher
redshifts, thereby providing additional constraining power, but also
due to the fact that the `gold' sample hints a small tension in the
ground-based `gold' sample data to prefer $j>1$ values~\footnote{An
  analysis of the `gold' sample data in which the HST supernovae are
  excluded leads to even stronger preference for $j>0$: $\Delta
  \chi^2_{\rm \mathcal{JQ}}=10.6$. In this case, for model
  $\mathcal{J}$ we obtain $q_{\rm 0}=-1.17\pm 0.28$ and
  $j=4.95^{+2.05}_{-1.84}$.}. Combining all three data sets, we obtain
a `strong' preference for model $\mathcal{J}$ over model
$\mathcal{Q}$, from all three statistical tests. Table~\ref{tab} shows
the mean marginalised parameters for each model and the $1\sigma$
confidence levels. Combining all three data sets, we obtain tight
constraints on $q_{\rm 0}=-0.81\pm 0.14$ and $j=2.16^{+0.81}_{-0.75}$.
Our result represents the first measurement of the jerk parameter from
cosmological data \footnote{Note that Riess et al. (2004) measured
  $j_{\rm 0}>0$ at the 2$\sigma$ level, where $j_{\rm 0}$ comes from a
  Taylor expansion of the Hubble parameter around small redshifts
  \citep{Visser:03}. As noted in subsection~\ref{previous} such an
  expansion is not appropiate when high redshift data are included, as
  in the `gold' sample.}.

Our dynamical analysis of the same three data sets gives
$w=-1.15^{+0.14}_{-0.18}$ and $\Omega_{\rm m}=0.306^{+0.042}_{-0.040}$
(see Table~\ref{tab:kidy}). Figure~\ref{dy_ki} shows the constraints
for both the kinematical $(q_{\rm 0},j$; top left panel) and dynamical
$(\Omega_{\rm m},w$; top right panel) models, using all three data
sets combined. In both cases, the dashed lines indicates the expected
range of results for $\Lambda$CDM models (i.e. a cosmological
constant). We find that both the kinematical and dynamical analyses of
the combined data are consistent with the $\Lambda$CDM model at about
the $1\sigma$ level.

It is important to recognise that the results from the kinematical and
dynamical analyses constrain different sets of departures from
$\Lambda$CDM. We are using two simple, but very different
parameterizations based on different underlying assumptions. The
results presented in Figure~\ref{dy_ki} therefore provide interesting
new support for the $\Lambda$CDM model.

The lower panels of figure~\ref{dy_ki} show the constraints obtained
for the three data sets when analysed individually.  It is important
to note the consistent results from the independent SNIa and X-ray 
cluster data sets.
Note that in the dynamical analysis, the
X-ray data provide valuable additional constraints on $\Omega_{\rm m}$,
when employing the $H_{\rm 0}$ and $\Omega_{\rm b}h^2$ priors.  The
overlap of all three data sets in both parameter spaces highlights the
robustness of the measurements. Comparing the upper and lower panels
of figure~\ref{dy_ki}, we see how the combination of data sets
significantly tightens the constraints.
 
\subsection{More complicated kinematical models}

For the combined data set, we have also searched for more complicated
departures from $\Lambda$CDM by including extra model parameters, as
described in Section~\ref{evoljerk}. We find no significant evidence
for models more complicated than a constant jerk model. In particular,
we find a negligible $\Delta \chi^2$ between models with constant jerk
$\mathcal{J}=[q_{\rm 0},j(c_{\rm 0})]$ and the next most sophisticated
model $\mathcal{J}_{\rm 1}=[q_{\rm 0},j(a;c_{\rm 0},c_{\rm 1})]$, and
between the latter model and the next one, $\mathcal{J}_{\rm
  2}=[q_{\rm 0},j(a;c_{\rm 0},c_{\rm 1},c_{\rm 2})]$.

\begin{figure}
\includegraphics[width=3.4in]{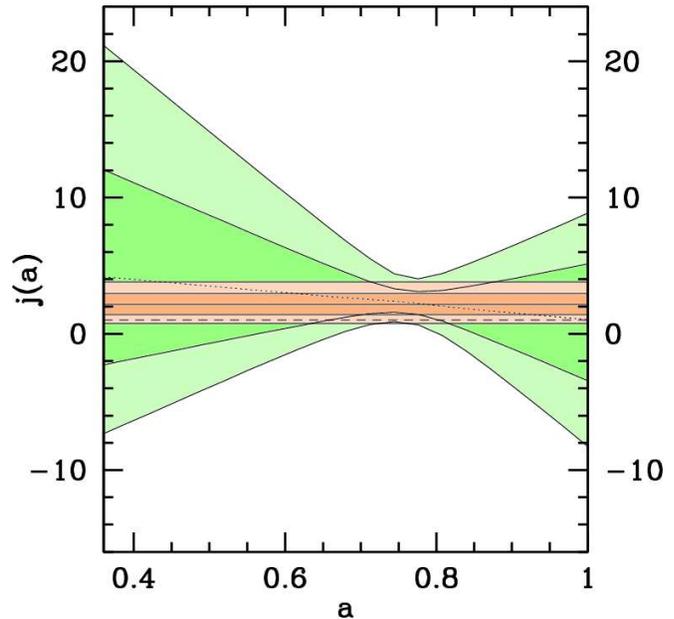}
\caption{The 68.3 and 95.4 per cent confidence variations about the
  median values for $j(a)$ as a function of the scale factor $a$, over
  the range where we have data [0.36,1]. Results are shown for the
  constant jerk model (model $\mathcal{J}$) (red, darker contours) and
  $\mathcal{J}_{\rm 1}$ model (green, lighter contours). In both
  cases, the constraints for all three data sets have been combined.
  The dashed line indicates the expectation, $j=1$ (constant) for a
  cosmological constant ($\Lambda$CDM) model.}
\label{co_j}
\end{figure}

It is, however, interesting to plot the differences between the
constraints obtained for each model. Figure~\ref{co_j} shows the
current $1\sigma$ and $2\sigma$ constraints around the median values
of $j(a)$ at different scale factors, $a$, over the range where we
have data $[0.36,1]$. The green, lighter contours show the constraints
for the $\mathcal{J}_{\rm_1}$ model and the red, darker contours for
the $\mathcal{J}$ model. From this figure it is clear that current
data provide the best constraints around $a\sim 0.77$, i.e. $z\sim
0.3$, and that at higher and lower redshift more data are required.
For the low redshift range, the forthcoming SDSS II SNIa data will be
helpful.  For the high redshift range, new HST SNIa and further X-ray
cluster data should be available in the near future. In the longer
term, SNIa data from the Large Synaptic Survey Telescope
(LSST)~\footnote{http://www.lsst.org/lsst$\_$home.shtml} and the
Supernovae Acceleration Probe (SNAP)~\footnote{http://snap.lbl.gov/},
and X-ray cluster data from
Constellation-X~\footnote{http://constellation.gsfc.nasa.gov/} should
provide tight constraints on both $j(a)$ and $w(a)$.  Future galaxy
redshift surveys covering a high redshift range will also help to
tighten these constraints, using the baryon oscillation experiment
\citep{Eisenstein:05,Cole:05}.

\subsection{Comparison of distance measurements }

\begin{figure}
\hspace{-0.7cm}
\includegraphics[width=3.3in,height=3in]{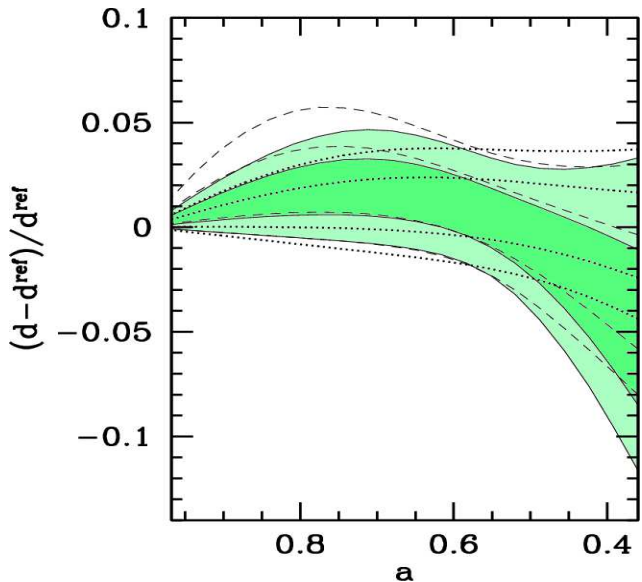}
\caption{The 68.3 and 95.4 per cent confidence limits on the offset in
  distance as a function of scale factor, relative to the reference
  $\Lambda$CDM cosmology, for both the kinematical (constant $j$;
  green, shaded curves) and dynamical (constant $w$; dotted and dashed
  curves) analyses.  The dotted curves show the results for the
  dynamical analysis in which the additional constraint on
  $\Omega_{\rm m}$ from the normalization of the $f_{\rm gas}$ curve
  is used. The dashed curve is for a dynamical analysis where this
  extra constraint on the normalization is ignored. The same MCMC
  samples used to construct Fig~\ref{dy_ki} have been
  used.}\label{fig:distance}
\end{figure}

It is interesting to compare directly the distance curves for the
kinematical (constant $j$) and dynamical (constant $w$) models, as
determined from the MCMC chains. Fig~\ref{fig:distance} shows the 68.3
and 95.4 per cent confidence limits on the offset in distance, as a
function of scale factor, relative to a reference $\Lambda$CDM
cosmology with $\Omega_{\rm m}=0.27$, $\Omega_{\Lambda}=0.73$. We see
that the kinematical and dynamical results occupy very similar, though
not identical, loci in the distance-scale factor plane. For the
dynamical analysis, the addition of the extra constraint on
$\Omega_{\rm m}$ from the normalization of the $f_{\rm gas}$ curve
tightens the constraints and pushes the results in a direction
slightly more consistent with the reference $\Lambda$CDM cosmology.

\subsection{Comparison with Riess et al. (2004)}

\begin{figure}
\hspace{0.3cm}
\includegraphics[width=2.9in]{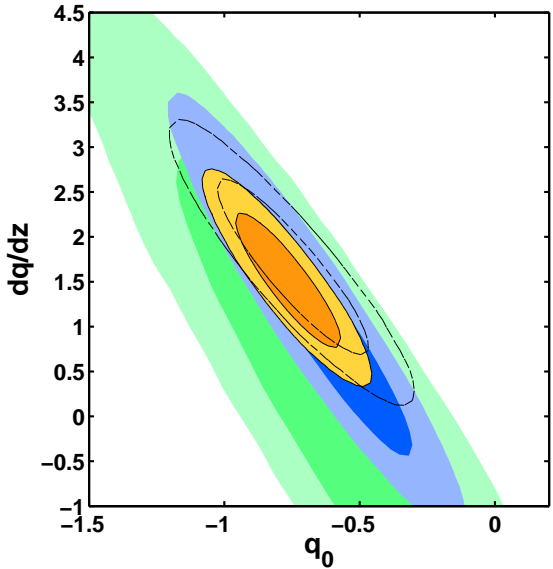}
\caption{The 68.3 and 95.4 per cent confidence limits in the ($q_0$,$dq/dz$) 
plane obtained using the SNIa data from the first year of the SNLS 
(Astier et al. 2005) (blue contours), the `gold' sample of Riess et al. (2004) 
(dashed contours), the cluster $f_{\rm gas}$ data of
Allen et al. (2006) (green contours) and the combination of all 
three data sets (orange contours).}
\label{deceleration_cons}
\end{figure}

For comparison purposes, we also present the results obtained using
the linear parameterization of $q(z)$ described by equation
(\ref{riess_dec}) and used by \cite{Riess04}.
Figure~\ref{deceleration_cons} shows the constraints in the plane
$(q_{\rm 0}, dq/dz)$ determined from each data set, and by combining
the three data sets (solid, orange contours).  It is clear that the
constraints from the three independent data sets overlap and that by
combining them we obtain significantly tighter results than using the
`gold' sample alone.

\section{The distance to the last scattering surface}
\label{cmbpoint}

Finally, we note that there is one further pseudo-distance measurement
available to us - the distance to the last scattering surface from CMB
data. Although this is not a purely kinematical data point, for
illustration purposes we show the constraints on $j(a)$ that can be
achieved if one is willing to make extra assumptions and include this
measurement. The extra assumptions involved, though strong, are
well-motivated. In detail, in order to use the distance to last
scattering, we assume that dark matter behaves like standard cold dark
matter at all redshifts, an assumption well tested by e.g. galaxy
cluster, weak lensing and galaxy redshift surveys at low redshifts and
CMB experiments at high-$z$. We also assume that pre-recombination
physics can be well described by a standard combination of cold dark
matter, a photon-baryon fluid and neutrinos, and that any early dark
energy component has a negligible affect on the dynamics. With these
assumptions, one can construct the comoving angular diameter distance
to the last scattering surface from $d_{\rm A}=r_{\rm
  s}(a_{dec})/\theta_{\rm A}$, where $r_{\rm s}(a_{rec})$ and
$\theta_{\rm A}$ are the comoving sound horizon at decoupling and the
characteristic angular scale of the acoustic peaks, respectively. For
a geometrically flat Universe with a negligible early dark energy
component, we calculate the sound horizon at decoupling as
\citep{Verde:03}

\begin{equation}
  r_{\rm s}(a_{\rm dec})\simeq \int^{a_{\rm dec}}_{0}\frac{c_{\rm s}(a)}
  {H_{\rm 0}(\Omega_{\rm m}a+\Omega_{\rm rad})^{1/2}}\,da\, ,
\label{sound}
\end{equation}

\noindent where $c_{\rm s}(a)=c/[1+(3\Omega_{\rm b}a)/(4\Omega_{\rm
  \gamma})]$ is the sound speed in the photon-baryon fluid,
$\Omega_{\rm rad}=\Omega_{\rm \gamma}+\Omega_{\rm \nu}$ is the present
radiation energy density, and $\Omega_{\rm \gamma}$ and $\Omega_{\rm
  \nu}$ are the present photon and neutrino energy densities,
respectively. We use our X-ray galaxy cluster data, assuming HST, BBNS
and $b$ priors, to determine $\Omega_{m}=0.27\pm0.04$ (Allen et al.
2006; note that this constraint mainly comes from low-redshift
clusters). We also use the COBE measurement of the CMB temperature
$T_{\rm 0}=2.725\pm0.002$K \citep{Mather:99} and a standard three
neutrino species model with negligible masses to obtain $\Omega_{\rm
  rad}$. For these constraints, we obtain $r_{\rm s}(z_{\rm
  dec})\simeq 146\pm 10$Mpc.

From \cite{Hinshaw:06} we have the multipole of the first acoustic
peak $l_{\rm 1}=220.7\pm 0.7$. This is related to $l_{\rm A}$ by a
shift $\phi$, $l_{\rm 1}=l_{\rm A}(1-\phi)$. Using the fitting formula
of \cite{Doran:01}, the BBNS prior for $\Omega_{\rm b}h^2$, a scalar
spectral index $n_{\rm s}=0.95\pm0.02$ \citep{Spergel:06} and assuming
no early dark energy, we find $\theta_{\rm A}=0.6\pm0.01$ degrees. We
then obtain a pseudo-model-independent distance to decoupling,
$d(z_{\rm dec})\simeq 13.8\pm1.1$Gpc, where $z_{\rm dec}=1088$
\citep{Spergel:06}.

\begin{figure}
\includegraphics[width=3.4in]{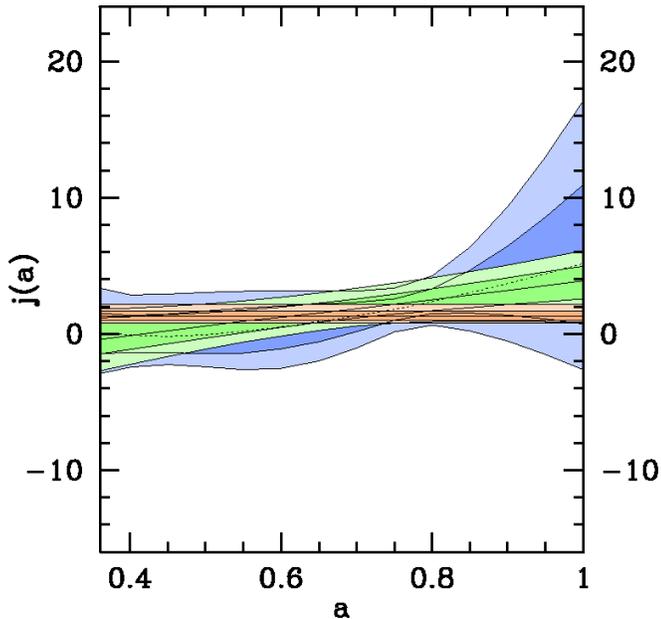}
\caption{$1\sigma$ and $2\sigma$ constraints on $j(a)$ over the range
  (including the distance to the last scattering surface) of the data
  [0.0009,1]. Note that this figure and figure~\ref{co_j} are plotted
  on the same scale for comparison purposes. This figure shows the
  same models as figure~\ref{co_j} plus the $\mathcal{J}_{\rm 2}$
  model, and uses the CMB prior as described at the text. The dotted
  line shows the median $j(a)$ curve for the $\mathcal{J}_{\rm 1}$
  model.}
\label{co_j_cmb}
\end{figure}

Fig~\ref{co_j_cmb} shows the tightening of the constraints obtained
using this ``data-point-prior'' at high redshift \footnote{Note that
  extending the analysis to the decoupling redshift $z_{\rm dec} =
  1088$ means that the radiation density becomes non-negligible.
  Although, $j$ can still be calculated as usual, $j^{\Lambda CDM}$
  will not equal 1 at these redshifts. However, the $\Lambda$CDM model
  can then be almost perfectly described as $j^{\Lambda
    CDM}(a)=1+2/(1+(a/a_{\rm eq}))$ \citep[see][for
  details]{Blandford:06} where $a_{\rm eq}$ is the mean marginalised
  scale factor at equality, from WMAP data. We have explicitly
  verified that, within the $1\sigma$ values of $a_{\rm eq}$,
  systematic offsets due to the affects of radiation have a negligible
  effect on the derived distances.}. Note that figure~\ref{co_j_cmb}
is plotted on the same scale as figure~\ref{co_j} and shows
$\mathcal{J}$ (red, darker contours) and $\mathcal{J}_{\rm 1}$ (green,
lighter contours) models as before, plus the $\mathcal{J}_{\rm 2}$
model (blue contours). Note also that here the range of the data is
$[a_{\rm min}=0.0009,a_{\rm max}=1]$. Again, using equation
(\ref{rescale}) we rescale the Chebyshev interval $[-1,1]$ to locate
the functions $\Delta j(a;\mathcal{C})$ in the range of scale factor
spanned by the data. The prior information at high redshift, from the
distance to last scattering, tightens the constraints significantly.
Evidently, the constraints from the kinematic analysis are sensitive
to the data quality at high redshift.

\section{Conclusions}
\label{discussion}

We have developed a new kinematical approach to study the expansion of
the history of the Universe, building on the earlier work of
\cite{Blandford:04}. Our technique uses the parameter space defined by
the current value of the cosmic deceleration parameter $q_{\rm 0}$ and
the jerk parameter $j$, where $q$ and $j$ are the dimensionless second
and third derivatives of the scale factor with respect to cosmic time.
The use of this $(q_{\rm 0},j)$ parameter space provides a natural
framework for kinematical studies. In particular, it provides a simple
prescription for searching for departures from $\Lambda$CDM, since the
complete set of $\Lambda$CDM models are characterized by $j=1$
(constant).

We have applied our technique to the three best available sets of
redshift-independent distance measurements, from type Ia supernovae
studies \citep{Riess04, Astier06} and measurements of the X-ray gas
mass fraction in X-ray luminous, dynamically relaxed galaxy clusters
\citep{Allen06}. Assuming geometric flatness, we measure $q_{\rm
  0}=-0.82\pm 0.14$ and $j=2.16^{+0.81}_{-0.75}$ (Figure~\ref{dy_ki}).
Note that this represents the first measurement of the cosmic jerk
parameter, $j$.  A more standard, dynamical analysis of the same data
gives $w=-1.15^{+0.14}_{-0.18}$ and $\Omega_{\rm
  m}=0.306^{+0.042}_{-0.040}$, also assuming flatness and HST, BBNS
and $b$ priors (Figure~\ref{dy_ki}). Both sets of results are
consistent with the standard $\Lambda$CDM paradigm, at about the
$1\sigma$ level.

In comparison to standard, dynamical approaches, our kinematical
framework provides a different set of simple models and involves fewer
assumptions. In particular, kinematical analyses such as that
presented here do not assume a particular gravity theory. The
combination of the kinematical and dynamical approaches therefore
provides important, complementary information for investigating late
time cosmic acceleration. The fact that both the kinematical and
dynamical results presented here are consistent with $\Lambda$CDM
provides important additional support for that model. The fact that
the two independent sets of distance measurements, from X-ray galaxy
clusters and supernovae, are individually consistent with
$\Lambda$CDM, is reassuring (Figure~\ref{dy_ki}).

We have searched for departures from $\Lambda$CDM using a new scheme
based on the introduction of Chebyshev polynomials. These orthonormal
functions allow us to expand any deviation from $\Lambda$CDM, $\Delta
j(a;\mathcal{C})$, as a linear combination of polynomials. We use the
coefficients of these polynomials, $\mathcal{C}$, as fit parameters.
The current data provide no evidence for a dependence of $j$ on $a$
more complicated than a constant value. However, higher order terms
may be required to describe future data sets. In that case, our scheme
has the advantage that, over a finite interval and using enough high
order terms, it will provide an acceptable global approximation to the
true underlying shape. Note that this scheme is also applicable to
dynamical studies of the evolution of the dark energy equation of
state, $w(a)$. Note also that Chebyshev polynomial expansions of the
same order for $w(a)$ and $j(a)$ explore a different set of models.
For example, a constant $j\neq 1$ model corresponds to an evolving
$w(a)$ model and vice versa.

We suggest that future studies should endeavour to use both
kinematical and dynamical approaches where possible, in order to
extract the most information from the data. The two approaches have
different strengths, can be applied to with a variety of data sets,
and are highly complementary. The combination of techniques may be
especially helpful in distinguishing an origin for cosmic acceleration
that lies with dark energy (i.e. a new energy component to the
Universe) from modifications to General Relativity.

\section*{Acknowledgment}

We acknowledge helpful discussions with A.~Frolov and technical
support from G.~Morris. The computational analysis was carried out
using the KIPAC XOC compute cluster at SLAC. SWA acknowledges
support from the National Aeronautics and Space Administration through
Chandra Award Number DD5-6031X issued by the Chandra X-ray Observatory
Center, which is operated by the Smithsonian Astrophysical Observatory
for and on behalf of the National Aeronautics and Space Administration
under contract NAS8-03060. RDB acknowledges support from National 
Science Foundation grant AST05-07732. This work was supported in part 
by the U.S. Department of Energy under contract number DE-AC02-76SF00515.

\bibliography{biblist_jerk}
\bibliographystyle{mn2e}


\label{lastpage}
\end{document}